\documentclass[12pt, oneside, a4paper]{article}
\usepackage{setspace}
\usepackage[a4paper,top=3cm,bottom=2cm,left=3cm,right=3cm,marginparwidth=1.75cm]{geometry}
\usepackage[english]{babel}
\usepackage[utf8x]{inputenc}
\usepackage[T1]{fontenc}
\usepackage{float}
\usepackage{siunitx}
\usepackage{mathtools} % http://ctan.org/pkg/mathtools
\usepackage{amsmath}
\usepackage{amssymb}
\usepackage{amstext}
\usepackage{graphicx}
\usepackage[colorlinks=true, allcolors=blue]{hyperref}
\usepackage{indentfirst}
\usepackage{calrsfs}
\usepackage{hyperref}
\usepackage{multicol}
\usepackage{listings}
\usepackage{cancel}
\usepackage{xcolor}
\usepackage{soul}

\newcommand{\hlf}{\frac{1}{2}}
\newcommand{\lt}{\left}
\newcommand{\rt}{\right}

\title{Benchmark: Tao's symplectic integration method}
\author{Matheus J. Lazarotto\textsuperscript{1,2}\footnote{Contact: matheus\_jean\_l@hotmail.com}, 
Iber\^{e} L. Caldas\textsuperscript{1} and
Yves Elskens\textsuperscript{2} \\
\footnotesize{1. Instituto de F\'{i}sica, Universidade de S\~{a}o Paulo, Rua do Mat\~{a}o 1371, S\~{a}o Paulo 05508-090, Brazil} \\
\footnotesize{2. Aix-Marseille Universit\'{e}, CNRS, UMR 7345 PIIM, F-13397, Marseille cedex 13, France} }

\date{\today}% It is always \today, today,
\begin{document}
\maketitle

\noindent
\textbf{Abstract:} A benchmark test was conducted for a new symplectic integration method originally 
developed by Molei Tao. The method raises interest due to its explicit evolution equation, with applicability 
to both separable and non-separable Hamiltonian systems, and an easy-to-implement, easily generalizable 
algorithm. In order to compare the method with other, more well-known methods, namely Störmer-Verlet 
and Runge-Kutta, we conducted a series of benchmark tests comparing their performance in terms of CPU 
time, system invariants functions conservation, and numerical symplectic area conservation. Overall, 
it was found that despite being considerably slower than the more optimized Runge-Kutta-Cash-Karp, Tao's method 
presents a similar performance to Störmer-Verlet, with the extra perk of being more generic and not 
requiring the use of implicit equations for the evolution of the equations of motion.

\section{Introduction}\label{sec:introduction}

Whenever one desires to numerically solve a system of differential equations with any integration method, 
the issue of precision in the solutions achieved is immediately raised. In classical mechanics, the 
solution of the Hamilton equations of motion is required in order to find the trajectories 
(orbits) over time from physical models of the most diverse sort. However, the Hamiltonian phase space
presents a topological structure such that some quantities are conserved, in particular the symplectic 
2-form and other adiabatic invariants \cite{Stuchi}. The symplectic 2-form is conserved regardless 
of the orbit being regular or chaotic, although numerically the latter is not achievable given its 
sensitive dependency on errors.

Perhaps the most widely known and disseminated methods for solving such equations are those of the 
Runge-Kutta (RK) family, including variations with fixed and adaptive time-step, such as Cash-Karp 
\cite{Cash-Karp}, Fehlberg \cite{Fehlberg}, and Dormand-Prince \cite{Dormand-Prince}. 
Although useful and remarkably fast, these methods lack the conservation of symplectic area, 
especially for long-time integrations. Thus, it is commonly required of an integration method 
to produce solutions that respect this conservation within numerical limits. 

For this purpose, the so-called symplectic methods were introduced, which keep a symplectic 2-form 
conserved with no divergent error by applying a canonical mapping between each discrete time step 
performed. The simplest of these methods, nonetheless still efficient, may be the St\"{o}rmer-Verlet 
one \cite{Hairer}. Despite many others being proposed, in general, they present harder implementation 
than those of the RK family, mostly due to their implicit mapping algorithm, thus requiring inversion 
or root-finding methods to solve at each step and therefore being slower in time performance.

The aim of this report is to divulge and test a new and practical symplectic integration method 
for both separable and non-separable Hamiltonian systems. The method was originally developed by 
Molei Tao and a thoroughly detailed description is given in its original reference \cite{Tao}. Here, 
the method will be briefly exposed, based on its original presentation, and its performance tested 
for a separable Hamiltonian systems, a 2D optical lattice model, and non-separable one, the restricted 
three-body planetary system. %, and the Walker-Ford Hamiltonian.

The interest in testing Tao's method is due to its explicit character and easy 
algorithm implementation, as opposed to the more commonly found symplectic algorithms that are implicit 
or exclusively explicit only for separable systems (see \cite{Tao} for references on previous explicit 
methods). As will be discussed further ahead, although the method does not aim for the fastest 
performance, at least when compared to traditional RK methods, it presented reasonable conservation 
of both energy and symplectic area for the 2D lattice system, and good symplectic conservation, but 
high energy deviations for the three-body model, making it an alternative option for those looking 
for an easy-to-use method when such features are required.

In the original exposition \cite{Tao}, the method's performance was compared to a 4th order RK  
and a previous pioneering explicit symplectic method developed by Pihajoki \cite{Pihajoki}, adding 
a correction for a binding problem, improving reliability for long integration times. Two 
numerical examples were chosen: the Schwarzschild geodesic problem and a non-integrable system, the 
turbulent nonlinear Schr\"{o}dinger equation, presenting better conservation of energy in both cases. 

Here further tests will be conducted on two other systems, the classical 2D optical lattice model and 
the restricted three-body problem.
% , and the Walker-Ford Hamiltonian. 
Similar to what was done in the original, a non-symplectic adaptive time-step Runge-Kutta-Cash-Karp (RKCK) 
and the symplectic St\"{o}rmer-Verlet methods will be used for comparison. The main tests presented here 
include different time steps, integration order of Tao's algorithm, binding parameters, and their effect 
on the conservation of invariant quantities of these systems.

Both systems were chosen for specific reasons, and brief introductions to each one are given in section 
\ref{sec:opt-lat} for the 2D lattice and section \ref{sec:three-body} for the restricted three-body problem. 
% and \ref{sec:walker-ford} for the Walker-Ford. 
For the lattice model, long integration times are commonly required in diffusion analysis considering 
orbits from a mixed phase space, thus presenting chaotic and regular solutions which need to be ascertained 
for their long-term conservation of constants of motion to ensure statistical precision. 
Similarly in the restricted three-body problem, long time integration is required to evaluate whether planets 
or satellites remain in orbit or leave indefinitely. 
% Analogously the Walker-Ford Hamiltonian presents mixed phase space and isochronous islands depending on 
% the selected resonances. 
Additionally, for the latter, it is a non-separable dynamical system, thus being able to test Tao's 
algorithm for one of its main application purpose.

In order to make this text self-contained, section \ref{sec:method} presents a summarized version of the 
algorithm completely based on the main reference \cite{Tao}, along with the tools used to evaluate the  
integration performance, regarding the conservation of invariants. Following the final remarks, an appendix 
shows the code implementation in C/C++ for generic (even) orders. A minimal example of how to use 
it is provided, along with the exact equations of motion used to allow easy reproduction of the tests 
made here (parameters and initial conditions are listed throughout the text).

The computational setup used was a Dell Vostro 3470 machine, with an Intel Core i7-8700 CPU (3.20 GHz) 
and Linux OS (Ubuntu 18.04), running simulations single-threaded and compiled with GCC++11. To enable 
complete reproduction of all the results, the original code used is made available at 
\url{github.com/matheuslazarotto/SymplecticTao}.

\section{Method}\label{sec:method}

\subsection{Tao's symplectic integration method}

To integrate a $d$-degree of freedom Hamiltonian system $H(\vec{q},\vec{p})$, with $\vec{q}$ and 
$\vec{p}$ as the $d$-dimensional position and momentum vectors respectively, Tao's method considers 
a new expanded Hamiltonian $\bar{H}$ including a new pair of copied variables $(\vec{q}_c,\vec{p}_c)$, 
starting from the same initial point ($\vec{q}_c(0) = \vec{q}(0)$, $\vec{p}_c(0)=\vec{p}(0)$) and 
evolving along with the main variables $\vec{q}$ and $\vec{p}$ from the equations of motion of the 
extended system:
\begin{equation}\label{eq:tao-hamiltonian}
	\bar{H}(\vec{q}, \vec{q}_c, \vec{p}, \vec{p}_c) = H_A + H_B + \omega H_C,
\end{equation}
where $H_A \coloneqq H(\vec{q},\vec{p}_c)$ and $H_B \coloneqq H(\vec{q}_c,\vec{p})$ are copies of 
the original Hamiltonian function written with interchanged variables; $\omega$ is a binding factor 
and $H_C$ is the coupling perturbation given by:
\begin{equation}
    H_C(\vec{q},\vec{q}_c,\vec{p},\vec{p}_c) \coloneqq \frac{||\vec{q}-\vec{q}_c||^2}{2} + \frac{||\vec{p}-\vec{p}_c||^2}{2}.    
\end{equation}

An integration step of $\delta t$ in Tao's method is thus made by a map $\phi_2^{\delta t}$ from 
Hamiltonian \ref{eq:tao-hamiltonian}, written as:
\begin{equation}\label{eq:lowest-mapping}
    \phi_{2}^{\delta t} \coloneqq \phi_{H_A}^{\delta t/2} \circ \phi_{H_B}^{\delta t/2} \circ \phi_{\omega H_C}^{\delta t} \circ \phi_{H_B}^{\delta t/2} \circ \phi_{H_A}^{\delta t/2},
\end{equation}
where each partial mapping is given by:
\begin{equation}
\begin{split}
    \phi_{H_A}^{\delta t} \coloneqq \begin{pmatrix} \vec{q} \\ \vec{p} \\ 
                                                    \vec{q}_c \\ \vec{p}_c \end{pmatrix} \to
                                    \begin{pmatrix} \vec{q} \\ \vec{p} - \delta t \nabla_{\vec{q}}H(\vec{q},\vec{p}_c) \\ 
                                                    \vec{q}_c + \delta t \nabla_{\vec{p}_c}H(\vec{q},\vec{p}_c) \\ \vec{p}_c \end{pmatrix},
                                    \quad\quad
    \phi_{H_B}^{\delta t} \coloneqq \begin{pmatrix} \vec{q} \\ \vec{p} \\
                                                    \vec{q}_c \\ \vec{p}_c \end{pmatrix} \to
                                    \begin{pmatrix} \vec{q} + \delta t \nabla_{\vec{p}}H(\vec{q}_c,\vec{p}) \\ \vec{p} \\ 
                                                    \vec{q}_c \\ \vec{p}_c - \delta t \nabla_{\vec{q}_c} H(\vec{q}_c, \vec{p}) \end{pmatrix},
\end{split}
\end{equation}
and
\begin{equation}
    \phi_{\omega H_C}^{\delta t} \coloneqq \begin{pmatrix} \vec{q} \\ \vec{p} \\ 
                                                           \vec{q}_c \\ \vec{p}_c \end{pmatrix} \to
                                       \hlf\begin{pmatrix} \begin{pmatrix} \vec{q}+\vec{q}_c \\ \vec{p}+\vec{p}_c \end{pmatrix} +
                                                R(\delta t)\begin{pmatrix} \vec{q}-\vec{q}_c \\ \vec{p}-\vec{p}_c \end{pmatrix} \\
                                                           \begin{pmatrix} \vec{q}+\vec{q}_c \\ \vec{p}+\vec{p}_c \end{pmatrix} -
                                                R(\delta t)\begin{pmatrix} \vec{q}-\vec{q}_c \\ \vec{p}-\vec{p}_c \end{pmatrix}
                                           \end{pmatrix},
                                        \quad\textrm{where}\quad
    R(\delta) \coloneqq \begin{pmatrix} \cos(2\omega\delta t) \mathbb{I}_d & \sin(2\omega\delta t) \mathbb{I}_d \\ 
                                       -\sin(2\omega\delta t) \mathbb{I}_d & \cos(2\omega\delta t) \mathbb{I}_d\end{pmatrix},
\end{equation}
with $\mathbb{I}_d$ as the $d$-dimensional identity matrix. As shown above, the method presents 
3rd-order error in $\delta t$, thus being a 2nd order method. However, higher order mappings are 
trivially obtainable by concatenating maps of lower order in a `triple jump' chain:
\begin{equation}\label{eq:triple-jump}
    \phi^\delta_l \coloneqq \phi^{\gamma_l\delta}_{l-2} \circ \phi^{(1-2\gamma_l)\delta}_{l-2} \circ \phi^{\gamma_l\delta}_{l-2},
\end{equation}
where $\gamma_l = (2 - 2^{\frac{1}{l+1}})^{-1}$ is a scaling factor. Note that the step size $(1-2\gamma)$ 
is negative, thus integrating movement backwards, even though the order of the step is very close to 
1. For the orders used here, we have $\gamma_6 = 1.12$ and $\gamma_4 = 1.17$, which do not significantly 
alter the step size magnitude. We point out that that higher orders must always be 
even $(l = 2n$, for $n \in \mathbb{Z}$), scaling down towards the lowest one.

As addressed by Tao, the binding factor $\omega$ must consider certain limits. That is, it must 
not be small enough in order to introduce numerical errors given the lack of bounding between the 
two copies that grow in time as $O(1)$ and nor too large. It is stated that $\delta t \ll 
\omega^{-\frac{1}{l}}$ ensures a good approximation for a feasible $\omega$.

Extended Hamiltonians were previously proposed for explicit methods for generic systems, particularly 
by Pihajoki \cite{Pihajoki}, from which Tao developed its own, adding the binding term $\omega H_C$ 
to correct the long term conservation of symplectic area. This correction bounds the error to order 
$O((\delta t)^l \omega^l)$, allowing much longer integration times.

\subsection{Performance evaluation}

To evaluate integration precision, two functions were calculated throughout time. For the lattice model, 
the Hamiltonian itself was used, which is an immediate constant of the motion and 
corresponds to the system's energy (section \ref{sec:opt-lat} %and \ref{sec:walker-ford} respectively 
for details), and for the three-body model, the Jacobi constant ($J = J(x,y,p_x,p_y)$ -- see section 
\ref{sec:three-body} for details). The other function is the symplectic non-degenerate 2-form $\delta S$ 
between two vectors $\vec{z} = (z^q_1, ..., z^q_d, z^p_1, ..., z^p_d)$ and $\vec{w} = (w^q_1, ..., 
w^q_d, w^p_1, ..., w^p_d)$ in a $d$-degrees of freedom phase space, given by:
\begin{equation}
    \delta S(\vec{z}, \vec{w}) \coloneqq \sum_{i=1}^{d} dq_i \wedge dp_i = \sum_{i=1}^{d} \lt(z_i^q w_i^p - z^i_p w_i^q\rt),
\end{equation}
which is expected to be conserved for a Hamiltonian flux, at least for integrable orbits when solved 
numerically \cite{Stuchi}.

For the calculation of the 2-form $\delta S$, an orbit for a given initial condition $\vec{s}_0$ is 
integrated along with a second one, slightly displaced from it with initial condition $\vec{s}_0 
\rightarrow \vec{s}_0 + \vec{\delta}$, for $|\vec{\delta}| = 10^{-10}$. The 2-form will thus measure 
the preservation of phase space areas projected into each plane ($q_i, p_i$) throughout time evolution. 
For chaotic solutions, an exponential divergence is expected between these orbits, but the longer it 
takes for them to diverge, the better. For integrable orbits, a good symplectic method is expected to 
keep the 2-form area limited and at the lowest possible value.

It is worth mentioning that symplecticity can be only numerically achieved for regular orbits, as 
guaranteed by the KAM theorem, since small perturbations do not disrupt invariant tori. On the other 
hand, in the case of chaotic orbits, this becomes utterly impossible due to their sensitive dependence 
on errors. Therefore, only the true solution of the equations of motion will preserve all the invariants 
of the system. Even though symplectic methods present an extra conservation of the non-degenerate 
2-form invariant, it may not account for all of them \cite{Stuchi}.

\section{2D Optical Lattice}\label{sec:opt-lat}

The first system under testing is a classical 2D optical lattice model described by the conservative 
separable Hamiltonian \cite{Horsley}:
\begin{equation}
    H(x, y, p_x, p_y) = p_x^2 + p_y^2 + U\lt(\cos^2(x) + \cos^2(y) + 2 \alpha \cos(x) \cos(y)\rt),
\end{equation}
with $p_x$, $p_y$ as canonical momenta, $U > 0$ an intensity parameter and $\alpha\in[0,1]$ a coupling 
parameter, such that when $\alpha=0$ the system is integrable. Units are normalized.

Optical lattices in general have been used experimentally to confine and control cold atoms for the 
study of new quantum states of matter \cite{Bloch1, Bloch2}. Theoretically, beyond the description 
of experiments, these Hamiltonian models are studied as both classical and quantum dynamical systems. 
In the classical realm particularly, they can be referred to as `soft billiards', relating to the 
hard walls billiard models, and are mainly used to simulate anomalous diffusion of particles through 
periodic potentials, thus requiring long integration times.

The lattice model considered here presents a mixed phase space for a wide range of its main control 
parameters, namely the energy $E=H(x,y,p_x,p_y)$ and the coupling $\alpha$. This allows one to fix 
a pair ($E$, $\alpha$) and select different initial conditions to study both chaotic and regular 
behaviors from the system's solutions. Selecting parameters $\alpha = 0.1$ and $E=25.0$, four orbits 
were chosen arbitrarily, two integrable and two chaotic, with the following initial points:
\begin{itemize}
    \item Trajectory 0 (Chaotic): {\color{blue}   $(x_0,y_0,p_{x0},p_{y0}) = (0.00000, 1.57070, -0.100000, 2.233745)$}
    \item Trajectory 1 (Regular): {\color{orange} $(x_0,y_0,p_{x0},p_{y0}) = (1.57070, 1.57070, -0.100000, 4.999000)$}
    \item Trajectory 2 (Regular): {\color{green}  $(x_0,y_0,p_{x0},p_{y0}) = (1.00000, 1.57070,  2.000000, 3.893746)$}
    \item Trajectory 3 (Chaotic): {\color{red}    $(x_0,y_0,p_{x0},p_{y0}) = (1.57070, 1.57070, -3.000000, 4.000000)$}
\end{itemize}

The points above were integrated for a total time of $t=300.0$, following the order of magnitude 
used in diffusion analysis. Since the system is spatially periodic, its position variables 
$(x,y)$ are modulated within the interval $[-\pi, \pi)$ by periodic boundary conditions. Therefore, 
the symplectic 2-form cannot surpass a maximum amount of order $10^0$, thus it does not diverge to 
infinity in the literal sense, rather it oscillates with this magnitude.

As mentioned in section \ref{sec:method}, the performance was evaluated by the deviation of the energy 
($\delta H$) from its initial value and the deviation of the symplectic 2-form ($\delta S$). All the data 
shown for the deviations comprise the average between the four orbits at the final time simulated. 
However, as expected, both chaotic orbits presented divergent behavior of the symplectic area after 
times $t \approx 10$; thus, all data regarding 2-form conservation are taken only over regular 
orbits. Section \ref{sec:opt-lat-dt-test} shows the algorithm performance conducted with different 
time steps and section \ref{sec:opt-lat-omega-test} tests made with different $\omega$ values.

\subsection{Time-step tests}\label{sec:opt-lat-dt-test}

In this section, we compare the performance of the different integration methods with Tao's method 
regarding different time-steps sizes and the conservation of invariants for the lattice model. The 
main results obtained are compiled in table \ref{tab:opt-lat-time-step}. For this test, all simulations 
using Tao's method used $\omega=500.0$, which yielded the best conservation results. Further tests with 
different $\omega$ are shown in section \ref{sec:opt-lat-omega-test}. Since the RKCK method has adaptive 
step size, its results were calculated for only one reference time step and placed in the $dt=10^{-6}$ 
column.

\begin{table}[H]
\centering
\begin{tabular}{c | c c c c | c c c c} \hline\hline
                       & \multicolumn{4}{c|}{$\delta H$}     & \multicolumn{4}{c}{$\delta S$} \\ \hline
          $dt$         & $10^{-3}$            & $10^{-4}$            & $10^{-5}$            & $10^{-6}$             & $10^{-3}$   & $10^{-4}$ & $10^{-5}$ & $10^{-6}$ \\ \hline
    RKCK               &        --            &       --             &         --           & $2\times10^{-10}$     &     --      &     --    &     --    & $10^{0}$  \\ \hline
    St\"{o}rmer-Verlet & $1\!\times\!10^{-4}$ & $2\!\times\!10^{-6}$ & $1\!\times\!10^{-8}$ & $1\!\times\!10^{-10}$ & $10^{-9}$   & $10^{-9}$ & $10^{-9}$ & $10^{-7}$ \\ \hline
    Tao ($l=2$)        & $6\!\times\!10^{0}$  & $1\!\times\!10^{-5}$ & $1\!\times\!10^{-7}$ & $2\!\times\!10^{-9}$  & $^*10^{1}$  & $10^{-7}$ & $10^{-7}$ & $10^{-7}$ \\ \hline
    Tao ($l=4$)        & $1\!\times\!10^{1}$  & $1\!\times\!10^{-5}$ & $9\!\times\!10^{-8}$ & $2\!\times\!10^{-8}$  & $^*10^{1}$  & $10^{-7}$ & $10^{-7}$ & $10^{-7}$ \\ \hline
    Tao ($l=6$)        & $8\!\times\!10^{0}$  & $1\!\times\!10^{-5}$ & $8\!\times\!10^{-8}$ & $3\!\times\!10^{-8}$  & $^*10^{1}$  & $10^{-7}$ & $10^{-7}$ & $10^{-7}$ \\ \hline
\end{tabular}
\caption{Energy ($\delta H$) and symplectic 2-form ($\delta S$) deviations for different integration 
         algorithms in different magnitude orders of time steps $dt$. The $^*$ mark indicates that 
         orbit 3 presented divergence whereas orbit 2 remained with $\delta S$ conserved.}
\label{tab:opt-lat-time-step}
\end{table}

One of the main features shown in table \ref{tab:opt-lat-time-step} is the lack of improvement 
in both energy and symplectic area in Tao's method when increasing the order $l$, particularly for 
$\delta S$, which remained constant at $10^{-7}$ (for $dt \leq 10^{-4}$). This same level of deviation  
was found in almost all simulations made, probably due to the fact that higher order mappings comprise 
only smaller steps of the order 2 map, halving the step size $dt$ for each order decrement. Therefore, 
up to order 6 this divides the original step size only by 4, thus not increasing its magnitude 
significantly. Nevertheless, as mentioned, this same level of conservation was found even 
for different binding factors, time step or order alike, possibly being related to a more fundamental 
reason.

In addition to the results shown in table \ref{tab:opt-lat-time-step}, an extension of the 
integration time was made to $t=3000$, for the cases $dt=10^{-4}$ and $dt=10^{-5}$. The purpose 
was to evaluate whether integration order delays the time required for energy deviation. However, 
no such behavior was found and energy deviations were bounded within the same magnitude for 
$dt=10^{-5}$ whereas for $dt=10^{-4}$ it diverged for times close to $t=1000$. Divergent runs did 
not present any pattern or delay of divergence time with the increment of integration order.

In general, besides the smallest time step used ($dt = 10^{-3}$), Tao's method presented similar 
conservation performance as St\"{o}rmer-Verlet. As expected, both symplectic methods present 
significantly lower performance in energy conservation magnitude when compared to RKCK, even though 
the latter presents secular growth and does not conserve symplectic area at all. The best performance 
found for symplectic methods had bounded energy fluctuations in the simulated time interval  
(exceptions will be mentioned in section \ref{sec:opt-lat-omega-test}).

The discrepancy between Verlet and Tao for $dt = 10^{-3}$ could be fixed with a different choice of $\omega$, 
since this parameter is directly related to the time-step selected. Even though the selected value of 
$\omega=500$ is well below the upper threshold $\delta t \ll \omega^{-\frac{1}{l}} = 0.35$ (for $l=6$), 
it may possibly be lower than a bottommost threshold.

In addition to conservation itself, the CPU time required for the simulations is also shown in table 
\ref{tab:opt-lat-time-step-CPU-time}. We point out that the computation time measured includes 
writing to file, symplectic 2-form and energy evaluations, although this process is common for all 
tests ran and for all methods, and may not interfere in the direct comparison between them. Furthermore, 
among the methods compared, Verlet and Tao were implemented manually with little to no optimization, 
whereas RKCK was used from the highly optimized GSL library \cite{GSL}, and being an adaptive step 
size method it has direct advantages from the other two.

\begin{table}[H]
\centering
\begin{tabular}{c c c c c} \hline\hline
            & \multicolumn{4}{c}{CPU time} \\ \hline
          $dt$         &  $10^{-3}$  &  $10^{-4}$  &  $10^{-5}$  &  $10^{-6}$  \\ \hline
    RKCK$^\dagger$     &      -      &      -      &       -     &      4sec   \\ \hline
    St\"{o}rmer-Verlet &     2sec    &    28sec    &  4min 57sec & 47min 24sec \\ \hline
    Tao ($l=2$)        &     3sec    &    29sec    &  5min 10sec & 50min 11sec \\ \hline
    Tao ($l=4$)        &     4sec    &    36sec    &  6min 1sec  & 59min 34sec \\ \hline
    Tao ($l=6$)        &     6sec    &    52sec    &  8min 43sec & 87min 47sec \\ \hline\hline
\end{tabular}
\caption{CPU time required for integration of the orbits for different methods and different 
         time steps. Parameters are the same used for the data in table \ref{tab:opt-lat-time-step}.}
\label{tab:opt-lat-time-step-CPU-time}
\end{table}

The tests conducted here should not be considered a rigorous benchmark, but rather as extra information 
providing the order of magnitude at which numerical algorithms may work. In particular, one may note 
that second-order Tao's method has a performance close to Verlet with similar CPU times, thus 
making it a good alternative. The slightly higher time in Tao's algorithm can be credited to the 
evaluation of sines and cosines, requiring considerable extra computation time when performed 
excessively along with the integration of the additional copied variables. Nevertheless, the method can 
be regarded as equivalent to the simple algorithm of Verlet, along with the extra feature of being 
generically applicable to non-separable systems.

\subsection{Binding factor tests}\label{sec:opt-lat-omega-test}

Analogously to the time step size tests previously shown, the performance of Tao's method for a series 
of binding parameter $\omega$ is shown in tables \ref{tab:opt-lat-omega-dt1e-4}, 
\ref{tab:opt-lat-omega-dt1e-5} and \ref{tab:opt-lat-omega-dt1e-6} for time steps of $dt=10^{-4}$, 
$dt=10^{-5}$ and $dt=10^{-6}$, respectively. As pointed out in section \ref{sec:method}, the choice 
of $dt$ is dependent on the value of $\omega$ and vice-versa, thus its testing requires a series of 
different magnitude values for both of them.

\begin{figure}[H]
    \centering
    \includegraphics[width=0.9\textwidth]{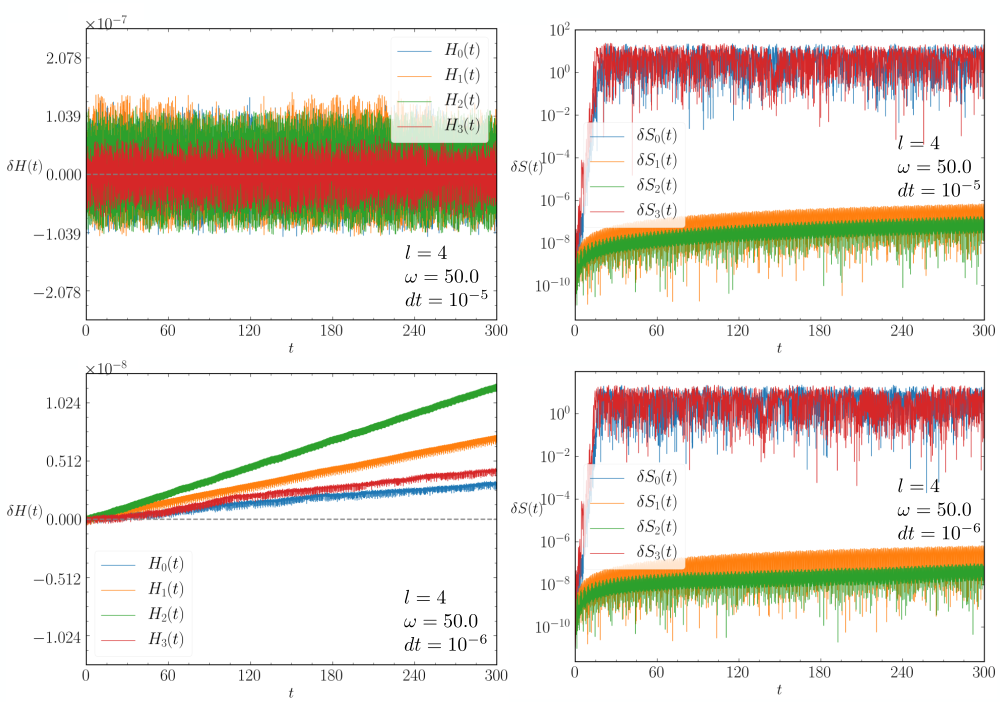}
    \caption{(Top row) Energy deviation $\delta H$ (left) and symplectic 2-form $\delta S$ (right) 
             along time for 4th order Tao's method with $\omega=50.0$ and $dt=10^{-5}$ depicting 
             limited growth and oscillatory fluctuations. (Bottom row) The same as the above but for 
             smaller time-step $dt=10^{-6}$ depicting secular growth of energy, at least up to the 
             time interval considered.}
    \label{fig:secular-growth}
\end{figure}

Besides the smaller time step used ($dt=10^{-4}$), which presented good conservation only for the 
middle range binding value ($\omega=500$), in all cases energy was conserved to at least order 
$10^{-7}$. However, for the smaller step size ($dt=10^{-6}$), several cases showed a clear linear 
growth of deviation for all orbits selected, the so-called secular deviation, even though its deviation 
did not surpass $10^{-7}$ over 300 time units. Figure \ref{fig:secular-growth} shows the comparison 
between a regular case with bounded energy and one with secular growth.

\begin{table}[H]
\centering
\begin{tabular}{c | c c c | c c c} \hline\hline
    $dt=10^{-4}$ & \multicolumn{3}{c|}{$\delta H$} & \multicolumn{3}{c}{$\delta S$} \\ \hline
    $\omega$     &  50  &  500  &  5000  &  50  &  500  &  5000 \\ \hline
    $l=2$        & $2\!\times\!10^{1}$ & $1\!\times\!10^{-5}$ & $1\!\times\!10^{1}$ & $1\!\times\!10^{-7}$ & $1\!\times\!10^{-7}$ & $1\!\times\!10^{-7}$ \\ \hline
    $l=4$        & $1\!\times\!10^{1}$ & $1\!\times\!10^{-5}$ & $9\!\times\!10^{0}$ & $1\!\times\!10^{-7}$ & $1\!\times\!10^{-7}$ & $1\!\times\!10^{-7}$ \\ \hline
    $l=6$        & $1\!\times\!10^{1}$ & $1\!\times\!10^{-5}$ & $9\!\times\!10^{0}$ & $1\!\times\!10^{-7}$ & $1\!\times\!10^{-7}$ & $1\!\times\!10^{-7}$ \\ \hline
\end{tabular}
\caption{Energy ($\delta H$) and symplectic 2-form ($\delta S$) deviations for different binding 
         values $\omega$ for Tao's method with time step $dt=10^{-4}$.}
\label{tab:opt-lat-omega-dt1e-4}
\end{table}

Regarding the conservation of the symplectic 2-form, for all cases, regular orbits presented the same 
deviation level, no greater than $10^{-7}$ regardless of integration order or 
binding value. Even in cases where energy deviation presented secular growth, $\delta S$ remained 
at the same conservation level. In most cases, an increase in $\omega$ does not improve energy 
conservation up to $t=300$ but it helped stop linear growth deviation. This indicates that 
some values may be too close to the bottom limit $\omega_0$, thus it is recommendable to take 
higher values, since these are safely below the upper limit of $dt \ll \omega^{-\frac{1}{l}} \approx 0.3$.

\begin{table}[H]
\centering
\begin{tabular}{c | c c c | c c c} \hline\hline
    $dt=10^{-5}$ & \multicolumn{3}{c|}{$\delta H$} & \multicolumn{3}{c}{$\delta S$} \\ \hline
    $\omega$     &  50  &  500  &  5000  &  50  &  500  &  5000 \\ \hline
    $l=2$        & $1\!\times\!10^{-7}$  & $1\!\times\!10^{-7}$ & $1\!\times\!10^{-7}$ & $1\!\times\!10^{-7}$ & $1\!\times\!10^{-7}$ & $1\!\times\!10^{-7}$ \\ \hline
    $l=4$        & $1\!\times\!10^{-7}$  & $1\!\times\!10^{-7}$ & $1\!\times\!10^{-7}$ & $1\!\times\!10^{-7}$ & $1\!\times\!10^{-7}$ & $1\!\times\!10^{-7}$ \\ \hline
    $l=6$        & $9\!\times\!10^{-8}$  & $1\!\times\!10^{-7}$ & $1\!\times\!10^{-7}$ & $1\!\times\!10^{-7}$ & $1\!\times\!10^{-7}$ & $1\!\times\!10^{-7}$ \\ \hline
\end{tabular}
\caption{Energy ($\delta H$) and symplectic 2-form ($\delta S$) deviations for different binding 
         values $\omega$ for Tao's method with time step $dt=10^{-5}$.}
\label{tab:opt-lat-omega-dt1e-5}
\end{table}

Unexpectedly, for $l=6$, the conservation became slightly worse as we increased $\omega$, although in 
all cases the limit $dt \ll \omega^{-\frac{1}{l}}$ was satisfied. For all $\omega$, the upper limit 
is given by $50^{-1/6}=0.52$, $500^{-1/6} = 0.3549$ and $5000^{-1/6}=0.2418$, thus making these 
results counter intuitive since as we increase $\omega$ we move away from the unspecified bottom 
limit $\omega_0$, being safely inside the `limits' for $dt$.

\begin{table}[H]
\centering
\begin{tabular}{c | c c c | c c c} \hline\hline
    $dt=10^{-6}$ & \multicolumn{3}{c|}{$\delta H$} & \multicolumn{3}{c}{$\delta S$} \\ \hline
    $\omega$     &  50  &  500  &  5000  &  50  &  500  &  5000 \\ \hline
    $l=2$        & $^*5\!\times\!10^{-9}$ & $^*3\!\times\!10^{-9}$ &   $2\!\times\!10^{-9}$ & $^*1\!\times\!10^{-7}$ & $^*1\!\times\!10^{-7}$ &   $1\!\times\!10^{-7}$ \\ \hline
    $l=4$        & $^*5\!\times\!10^{-9}$ & $^*3\!\times\!10^{-9}$ & $^*2\!\times\!10^{-9}$ & $^*1\!\times\!10^{-7}$ & $^*1\!\times\!10^{-7}$ & $^*1\!\times\!10^{-7}$ \\ \hline
    $l=6$        & $^*4\!\times\!10^{-9}$ &   $3\!\times\!10^{-8}$ &   $1\!\times\!10^{-8}$ &   $1\!\times\!10^{-7}$ &   $1\!\times\!10^{-7}$ &   $1\!\times\!10^{-7}$ \\ \hline
\end{tabular}
\caption{Energy ($\delta H$) and symplectic 2-form ($\delta S$) deviations for different binding 
         values $\omega$ for Tao's method with time step $dt=10^{-6}$. Fields marked by $^*$ presented 
         secular growth of energy for all cases.}
\label{tab:opt-lat-omega-dt1e-6}
\end{table}

In the previous section, all data exposed was made for $\omega=500$, since this parameter yielded 
the best result considering all time steps. Notwithstanding, the best results for energy conservation 
were obtained for $\omega=5000$ and $dt=10^{-6}$, however not for arbitrary orders. In general, changes 
in the binding factor did not alter much the order of conservation for $dt=10^{-5}$ and $dt=10^{-6}$, 
but implied secular growth in the latter. Higher sensitivity was found for $dt=10^{-4}$.

\pagebreak

\section{Restricted three-body problem}\label{sec:three-body}

\begin{figure}[H]
\begin{minipage}[c]{0.59\textwidth}
    \centering
    \includegraphics[width=\textwidth]{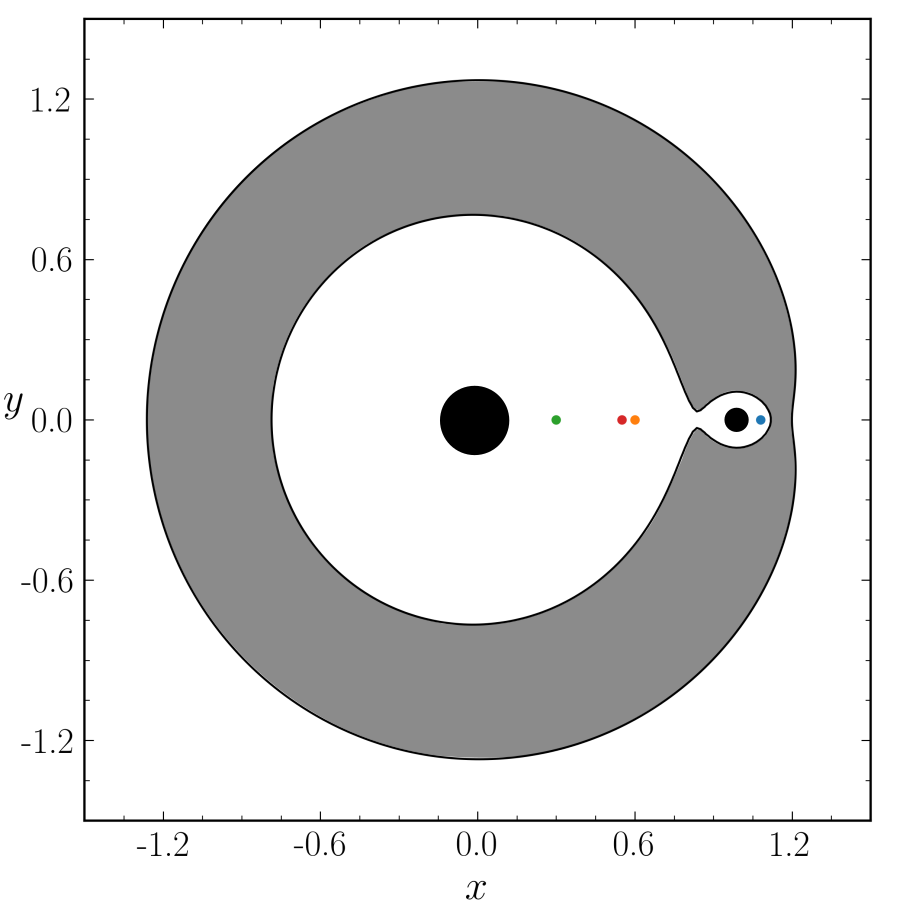}
\end{minipage}\hfill
\begin{minipage}[c]{0.40\textwidth}
    \caption{Spatial plot of the planar restricted three-body model scheme. Black circles indicate the massive 
             bodies and the colored dots the initial conditions used. The gray area corresponds to 
             Hill's region, where the probe particle is not allowed to enter due to the restriction 
             $2\Omega(x,y) > J$, bounded by the zero-velocity curve. For the parameters selected, 
             ($J=3.1843, \mu_1=0.9879, \mu_2=0.0121$), all orbits are enclosed and cannot escape the 
             surroundings of the massive bodies, being allowed to move only within the inner white area. 
             It can orbit either the heavier or the lighter mass or transit between the two.}
    \label{fig:three-body-scheme}
\end{minipage}
\end{figure}

One of the simplest examples of $n$-body planetary systems, yet still presenting complex solutions, 
is given by the restricted three-body problem. This classical formulation considers two 
heavy masses and a probe particle with negligible mass compared to the more massive ones, moving 
along the gravitational potential of the two massive bodies (figure \ref{fig:three-body-scheme}). 
This scenario can possibly represent a triad of a star, a planet and a small moon, or a planet, its 
moon and a satellite.

The most common representation of the system is from a rotational frame of reference, with the heavy 
bodies fixed along the $y=0$ axis, 1 space unity apart from each other. This relates to two 
main aspects of the model. First, that the heavy bodies orbit around its center of mass with the same 
angular velocity $n$ (which is also the velocity of the rotating frame), and second, that a unitary 
lenght scale is settled. In this setup, the masses are scaled so that the reduced mass is unitary, 
\textit{i.e.}, $\mu = G(\mu_1 + \mu_2) = 1$, where $G$ is the gravitational constant and 
$\mu_i = m_i / \lt(m_1 + m_2\rt)$ is the reduced mass of the massive bodies \cite{Murray, Oliveira}.

The probe dynamics is thus restricted to be coplanar with the massive bodies and its movement dictated 
by a potential $\Omega(x,y)$, resulting from the superposition of the individual gravitational interactions 
with the two heavy bodies and an term accounting for the centrifugal force due to the 
rotating frame:
\begin{equation}\label{eq:omega}
    \Omega(x,y) = \frac{n}{2}\lt(x^2 + y^2\rt) + \frac{\mu_1}{\sqrt{(x + \mu_2)^2 + y^2}} + \frac{\mu_2}{\sqrt{(x - \mu_1)^2 + y^2}}
\end{equation}

In the Hamiltonian formulation, one can write the particle's canonical variables as its position and a 
pair of conjugated momenta ($p_x, p_y$), and the system Hamiltonian written as
\begin{equation}\label{eq:three-body-hamiltonian}
    H(x,y,p_x,p_y) = -2 J(x,y,p_x,p_y)
\end{equation}
where
\begin{equation}\label{eq:jacobi-constant}
    J(x,y,p_x,p_y) = 2 \Omega(x,y) - (p_x + y)^2 - (p_y - x)^2
\end{equation}

It is worth mentioning that since the coordinate system $(x,y)$ is rotating, the Hamiltonian function 
\ref{eq:three-body-hamiltonian} does not correspond to the particle's energy. Instead, it corresponds 
to the so called Jacobi function $J$, which is itself a constant of motion.

Analogously to the procedure shown in section \ref{sec:opt-lat} for the lattice model, 
we selected the system parameters with a Jacobi constant of $J=3.1843$ and reduced masses $\mu_1 
= 0.9879$ and $\mu_2 = 1 - \mu_1 = 0.0121$. In accordance with these parameters, four initial 
conditions were selected:
\begin{itemize}
    \item Trajectory 0 (Chaotic): {\color{blue}   $(x_0,y_0,p_{x0},p_{y0}) = (1.080000, 0.000000, -0.080000, 1.300003)$}
    \item Trajectory 1 (Regular): {\color{orange} $(x_0,y_0,p_{x0},p_{y0}) = (0.600000, 0.000000,  0.000000, 1.282517)$}
    \item Trajectory 2 (Regular): {\color{green}  $(x_0,y_0,p_{x0},p_{y0}) = (0.300000, 0.000000,  0.000000, 2.108413)$}
    \item Trajectory 3 (Chaotic): {\color{red}    $(x_0,y_0,p_{x0},p_{y0}) = (0.550000, 0.000000,  0.000000, 1.246951)$}
\end{itemize}

A total time of $t=1000.0$ was used for the integration. The initial conditions were selected such that 
two of them are chaotic and two regular. None of them collide with either of the massive bodies, in 1
the sense that their distance to any of the divergence points does not get smaller than $10^{-6}$. Thus, 
integration can be performed without the necessity of renormalization techniques.

% The same tests conducted here were made over the velocity representation of the restricted three-body 
% problem and no difference was seen, with the symplectic 2-form being equally conserved, even though the 
% velocities are not the canonical pair of variables for this system.

\subsection{Performance tests}\label{sec:three-body-dt-test}

Similar to what was done for the lattice model in section \ref{sec:opt-lat}, this section presents 
the performance comparison between different integration methods and Tao's symplectic one for a series 
of its parameters. Different magnitudes of time step, binding factor $\omega$, and integration 
order were considered for Tao's method. As before, integration performance was evaluated by the 
deviations of symplectic 2-form ($\delta S$) and Jacobi constant ($\delta J$), which is a constant of 
motion for the restricted three-body problem \cite{Oliveira}. As will be discussed shortly, Tao's 
second-order ($l=2$) yielded the best performances, thus, for simplicity, results for higher 
orders were suppressed.

\begin{figure}[H]
    \centering
    \includegraphics[width=1.0\textwidth]{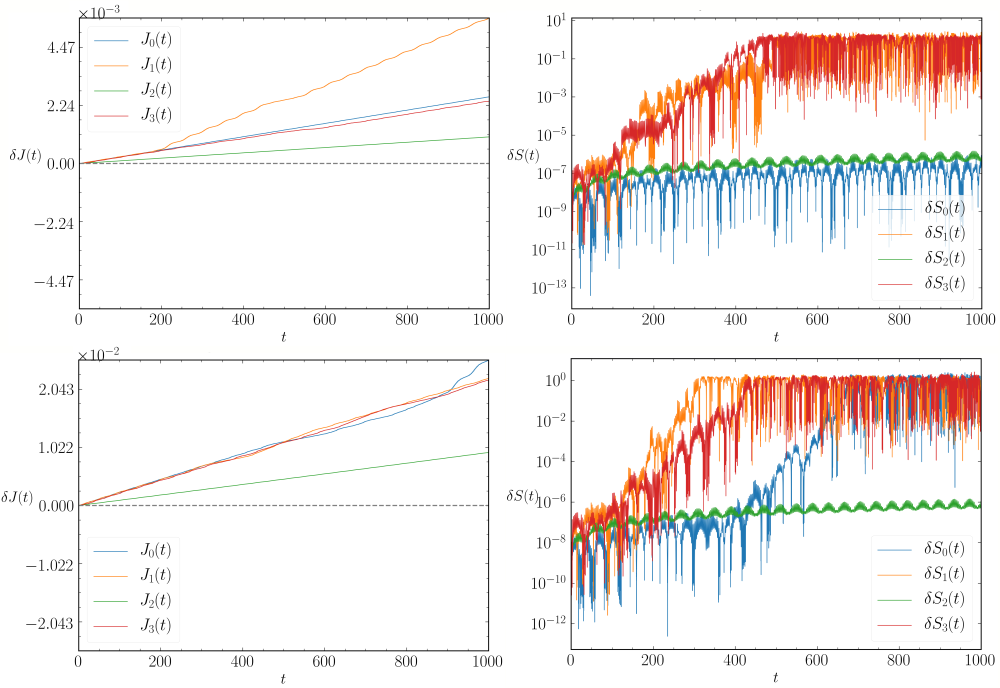}
    \caption{(Top row) Jacobi constant $\delta J$ (left panel) and symplectic 2-form $\delta S$ 
             (right panel) deviations for the four orbits simulated for the best performance found 
             (parameters: $l=2$, $\omega=5\times10^{6}$, $dt=10^{-6}$).
             (Bottom row) The same as the top row but for a case presenting deviation (parameters:
             $l=6$, $\omega=5\times10^{4}$, $dt=10^{-6}$).}
    \label{fig:three-body-plot}
\end{figure}

The results were divided by the Jacobi constant, as shown in table \ref{tab:three-body-jacobi-constant}, 
and symplectic 2-form, shown in table \ref{tab:three-body-symplectic-area}. All data displayed for 
Jacobi constant comprises the average of all four orbits simulated, whereas those for the 
symplectic 2-form consider only the two regular orbits. Cases in which divergence occurred are marked 
by an asterisk (figure \ref{fig:three-body-plot}). For the two other methods used for comparison, 
St\"{o}rmer-Verlet and RKCK, since they have no relation with a binding factor, their results 
are placed in a single column in order to be displayed along with Tao's results.

As a general trend, both symplectic methods presented poor conservation of the Jacobi 
constant, not being better than $10^{-3}$ (see table \ref{tab:three-body-jacobi-constant}). Even though 
St\"{o}rmer-Verlet is not applicable to non-separable systems, it was used here to 
compare with the poor performance found by Tao's method. For comparison, RKCK yielded a deviation of $10^{-14}$. 
Moreover, in the lattice model previously discussed, the same comparison between 
symplectic and RKCK methods showed the difference in magnitude of two orders. No detailed investigation 
was made on this difference between the two systems, however, it seems related to the non-separability 
of the three-body system motion equations. Although the method is supposed to handle this kind of 
scenario, it presented a similar behavior then St\"{o}rmer-Verlet which is not expected to.

Regarding the variation of the binding factor $\omega$, the magnitudes selected for testing here showed 
the best results. Nevertheless, a more extensive investigation for values within a narrow range of 
the orders of magnitude considered here may possibly yield better results. Even though no significant 
variation was found for the binding factor for a given time-step, this shows that depending on the 
integration order, it can disrupt conservation. For the selected binding factors, the lower estimated 
upper limit for the time-step ($\delta t \ll \omega^{-\frac{1}{l}}$) is for $l=2$ and $\omega=5\times10^{6}$, 
and it is given by $4.4\times10^{-4}$, thus still one order of magnitude higher than the time-step 
used.

Given the low performance, larger time-steps were not considered since no increment is expected for 
these cases, particularly since a higher magnitude approaches the upper limit of the time-step. 
For lower magnitudes, although it shall improve conservation, it were not tested due to its long 
computational cost for the symplectic algorithms used here. For the magnitudes selected, $dt=10^{-5}$ 
and $dt=10^{-6}$, an expected improvement of one order of magnitude is achieved with the smaller one. 
However, as seen for the lattice model, no significant enhancement is seen for higher integration 
orders.

\begin{table}[H]
\centering
\begin{tabular}{c | c c c c c c} \hline\hline
                       & \multicolumn{6}{c}{$\delta J$} \\ \hline
          $dt$         & \multicolumn{3}{c|}{$10^{-5}$}       & \multicolumn{3}{c}{$10^{-6}$} \\ \hline
        $\omega$       & \multicolumn{1}{c|}{$5\times10^{4}$} & \multicolumn{1}{c|}{$5\times10^{5}$} & \multicolumn{1}{c|}{$5\times10^{6}$} & \multicolumn{1}{c|}{$5\times10^{4}$} & \multicolumn{1}{c|}{$5\times10^{5}$} & \multicolumn{1}{c}{$5\times10^{6}$} \\ \hline
          RKCK         &        --          &         --         &        --          &         --         &        --          & $10^{-14}$         \\ \hline
    St\"{o}rmer-Verlet &        --          &         --         & $4\times10^{-2}$   &         --         &        --          & $3\times10^{-3}$   \\ \hline
    Tao ($l=2$)        & $2\times10^{-2}$   & $2\times10^{-2}$   & $2\times10^{-2}$   & $2.6\times10^{-3}$ & $2.8\times10^{-3}$ & $2.6\times10^{-3}$ \\ \hline
    % Tao ($l=4$)        & $5\times10^{-2}$   & $5\times10^{-2}$   & $5\times10^{-2}$   & $6.0\times10^{-3}$ & $6.4\times10^{-3}$ & $6.0\times10^{-3}$ \\ \hline
    % Tao ($l=6$)        & $1.6\times10^{-1}$ & $1.6\times10^{-1}$ & $1.6\times10^{-1}$ & $2.0\times10^{-2}$ & $1.7\times10^{-2}$ & $1.7\times10^{-2}$ \\ \hline
\end{tabular}
\caption{Jacobi constant deviations ($\delta J$) for different integration algorithms for different 
         magnitudes of time step $dt$ and binding factor $\omega$ (regarding Tao's method only). }
        %  The $^*$ mark indicates that orbit 3 presented divergence whereas orbit 2 remained with 
        %  $\delta S$ conserved.}
\label{tab:three-body-jacobi-constant}
\end{table}

All tested cases showed secular deviation of the energy (as illustrated in figure 
\ref{fig:three-body-plot}), with no bounded oscillations as seen for the 2D lattice model, even though 
some of them still showed logarithmic deviation of the symplectic area. However, divergence of symplectic 
2-form was still commonly found. Particularly for $dt=10^{-5}$, all cases showed divergence of the 
symplectic area for at least one of the regular orbits.

Similar to the results on the lattice model, conservation of symplectic area did not present any 
improvement in magnitude, remaining at a constant level of $10^{-7}$ regardless of variation of 
$\omega$, $dt$ or $l$. The main difference seems to affect the conservation itself, which shows 
complete divergence.

\begin{table}[H]
\centering
\begin{tabular}{c | c c c c c c} \hline\hline
                       & \multicolumn{6}{c}{$\delta S$} \\ \hline
          $dt$         & \multicolumn{3}{c|}{$10^{-5}$}       & \multicolumn{3}{c}{$10^{-6}$} \\ \hline
        $\omega$       & \multicolumn{1}{c|}{$5\times10^{4}$} & \multicolumn{1}{c|}{$5\times10^{5}$} & \multicolumn{1}{c|}{$5\times10^{6}$} & \multicolumn{1}{c|}{$5\times10^{4}$} & \multicolumn{1}{c|}{$5\times10^{5}$} & \multicolumn{1}{c}{$5\times10^{6}$} \\ \hline
          RKCK         &    --  &  --  &     --        &        --     &      --      & $10^{-1}$                   \\ \hline
    St\"{o}rmer-Verlet &    --  &  --  &  $10^{-6}$    &        --     &      --      & $10^{-7}$                   \\ \hline
    Tao ($l=2$)        & $10^{-7}(*)$  & $10^{-7}(*)$  & $10^{-7}(*)$  & $10^{-7}$    & $10^{-7}$    & $10^{-7}$    \\ \hline
    % Tao ($l=4$)        & $10^{-7}(*)$  & $10^{-7}(*)$  & $10^{-7}(*)$  & $10^{-7}$    & $10^{-7}$    & $10^{-7}$    \\ \hline
    % Tao ($l=6$)        & $10^{-6}(**)$ & $10^{-6}(**)$ & $10^{-6}(**)$ & $10^{-7}(*)$ & $10^{-7}(*)$ & $10^{-7}(*)$ \\ \hline
\end{tabular}
\caption{Symplectic 2-form deviations ($\delta S$) for different integration algorithms for different 
         magnitudes of time steps $dt$ and binding factor $\omega$ (regarding Tao's method only). 
         Cases marked with $(*)$ presented secular divergence on one of the regular orbits.}
     %   and $(**)$ this deviation was found for both regular orbits.}
\label{tab:three-body-symplectic-area}
\end{table}
\pagebreak

\section{Final Remarks}

The series of tests performed here presented diverse results for the two systems selected: the 2D 
lattice classical model and the restricted three-body problem. For the 2D lattice model, good 
performance was achieved for some parameters of Tao's algorithm (time-step, binding factor, and 
integration order), with bounded energy deviations of the order of $10^{-9}$ and the symplectic 2-form 
preserved at the order of $10^{-7}$. The results for the lowest order ($l=2$) were comparable in both 
performance and CPU time to the St\"{o}rmer-Verlet. When compared to RKCK, an expectedly lower energy 
conservation was achieved but still at an acceptable level.

For the three-body problem, poor performance in the conservation of the Jacobi constant was found 
for all binding factors, time-step and integration order values considered. Even though symplectic 2-form 
was kept conserved to order $10^{-7}$, many cases still diverged along with Jacobi's constant, 
which did not present deviations smaller than order $10^{-3}$. For the tests on this system 
regarding the binding factor $\omega$, no bounded energy preservation was found, raising the question 
of whether a narrow interval of $\omega$ values, within the magnitude values selected here, could yield a 
better result then the ones obtained.

In general for both systems, higher orders of Tao's method, up to the sixth, did not improve 
conservation of either the Hamiltonian constants (energy and Jacobi constant) or the symplectic 2-form. 
Moreover, this also implies that the construction of an adaptive time step version of the algorithm 
may not be trivially achieved by comparing step sizes of different precision for error evaluation. 

Given the higher dependency of performance with the binding factor chosen, the use of Tao's method 
requires special attention to the binding factor. As shown in section \ref{sec:opt-lat-omega-test} 
for the lattice system, values that are too low can easily lead to secular energy deviation or even 
complete lack of conservation. For this system, with step sizes of order $dt=10^{-5}$, a large range 
of values covering at least two orders of magnitude ($\omega > 50$) showed good conservation results.

Even though Tao's method does not aim at speed, but rather aims at symplecticity and ease of implementation 
for generic Hamiltonian systems, its CPU time is drastically slower when compared to the RKCK method. 
We recommend its use for integrating single orbits that require the conservation of symplectic area. 
However, further benchmarks and code optimization could provide better information and the possibility 
of improved performance and calculation of averages over full phase spaces, for example.
We call attention to the binding factor choice, since different initial conditions may require considerably 
different values, possibly with complete lack of conservation.

\section*{Acknowledgments}

M. Lazarotto would like to acknowledge Vitor M. Oliveira for valuable discussions on symplectic 
integration and on the restricted three-body problem. The authors also acknowledge 
the financial support from the Brazilian federal agencies CNPq, grants 302665/2017-0 and 141750/2019-7 
and the S\~{a}o Paulo Research Foundation (FAPESP, Brazil), grant 2018/03211-6.

\bibliographystyle{unsrt}
\bibliography{ref.bib}

\section*{Appendix}
A code implementation made in C/C++ programming language is shown below, along with a minimal example 
on how to use it. The simplicity of the method's algorithm makes it easy to port to other languages, 
requiring basic computational features. The dynamical equations used for the 2D optical lattice and 
the three-body model are also available at the final subsection, allowing for complete reproduction 
of the tests conducted here. 
The full code can be retrieved at: \url{https://github.com/matheuslazarotto/SymplecticTao}.

\subsection*{Minimal usable example}
A minimal setup example to integrate an orbit from a initial condition \texttt{x0, y0, px0, py0} 
for a total time \texttt{run\_time} is presented here. It starts by showing the system's dynamical 
functions defined in \texttt{system\_functions()}, allowing the use of external arguments, 
\textit{i.e.}, constants or parameters of any sort stored in an array \texttt{*args}. Then, 
\texttt{main()} function is where integration is ran up to \texttt{run\_time}, in time steps of 
size \texttt{dt}, while evolving the array with the dynamical \texttt{variables} (and in parallel the copy 
variables \texttt{variables\_copy}), along with other parameters required by the method, namely 
the \texttt{order}, \texttt{omega} and \texttt{gamma}.

This example considers a generic conservative and separable Hamiltonian function with two degrees 
of freedom (therefore a 4D dynamical system), with cartesian coordinates (\texttt{x,y,px,py}) 
and motion equations given by the functions: \texttt{dqxdt(), dqydt(), dpxdt(), dpydt()}. Although, 
this implementation is trivially generalizable to any number of degrees of freedom. 
This pictoric example assumes two constant 
parameters (\texttt{k} and \texttt{m}), stored in the array \texttt{args[] = \{k, m\}}. 

The explicit code for Tao's method function step \texttt{ode\_step\_symplectic()} is displayed 
further in this appendix.

\begin{verbatim}

int system_functions(double t, const double var[], double functions[], void *args) 
{
    (void)(t); // Avoid unused parameter warning
    double *par = (double *)args;
    double k = par[0];
    double m = par[1];
    
    f[0] = dqxdt(var[2], m);
    f[1] = dqydt(var[3], m);
    f[2] = dpxdt(var[0], var[1], k);
    f[3] = dpydt(var[1], var[0], k);
    
    return 0;
}

int main()
{
    const int system_dimension = 4;
    const double run_time = 100.0;
    const double x0 = 1.0;
    const double y0 = 1.0;
    const double px0 = 5.0;
    const double py0 = 1.0;
    const double k = 1.0;
    const double m = 1.0;
    const double args[] = {k, m};
    const double dt = 1e-4;
    double time = 0.0;
    double variables = {x0, y0, px0, py0};
    double variables_copy[] = {x0, y0, px0, py0};
    const int order = 4;                   
    const double omega = 100.0;
    const double gamma = 1.0 / (2.0 - pow(2.0, 1 / (order + 1)));
    const bool lowest = (order == 2) ? true : false;

    /* Integration loop */
    while (t <= run_time)
    {
        ode_step_symplectic(variable, variable_copy, t, dt, 
                            omega, gamma, system_dimension, 
                            order, lowest, args, system_functions);
    }
}
\end{verbatim}

\subsection*{Method implementation}

Below is the implementation of the functions required to run the integration algorithm. 
The main function used is \texttt{ode\_step\_symplectic()}, which evolves the system 
\texttt{variables} (and its copy in parallel) for one time step \texttt{dt}, for a given integration 
order, binding rotation \texttt{omega}, and time \texttt{gamma} parameters.

The \texttt{ode\_step\_symplectic()} function calls itself recursively when applying `triple jumps', 
(see section \ref{sec:method}), diminishing to lower orders (in even steps) until the lowest 
one (2) is reached, where the last `triple jump' chain is called as \texttt{ode\_step\_symplectic\_lowest()}. 
However, if one applies the method with order 2, it is required to be different of when the lowest case 
(order = 2 ) is reached from the downscalling when recursively called from a higher order method. 
When one simply wants to integrate using order 2, no triple jump is required, only the regular mapping (eq. \ref{eq:lowest-mapping}), 
whereas when reaching order 2 by downscalling, a triple jump of order 2 $\phi$ mappings must be called
(eq. \ref{eq:triple-jump}). For this purpose, a boolean flag \texttt{lowest} is set to \texttt{true} in 
case only direct order 2 integration is made, or \texttt{false} when higher order downscalling must be 
used. The partial mappings $\phi_A$, $\phi_B$ and $\phi_C$ are separately implemented.

\begin{verbatim}
void ode_step_symplectic(double var[], double var_copy[], double &t, double dt, 
                         double w, double gamma, unsigned int system_dim, int order, 
                         bool lowest, double *args, int (*system_function)(double t, 
                         const double s[], double f[], void *args))
{
    /* Recursively call `triple jumps' of Phi maps down to the lowest order (2) */
    if (lowest)
    {
        ode_step_symplectic_lowest(var, var_copy, t, dt, w, system_dim, 
                                   args, system_function);
    }
    else
    {
        order -= 2;
        double dt_s = gamma * dt;
        double dt_m = (1.0 - 2.0 * gamma) * dt;
        if (order > 2)
        {
            double local_gamma = 1.0 / (2.0 - pow(2.0, 1 / (order + 1)));
            ode_step_symplectic(var, var_copy, t, dt_s, w, local_gamma, system_dim, 
                                order, lowest, args, system_function);
            ode_step_symplectic(var, var_copy, t, dt_m, w, local_gamma, system_dim, 
                                order, lowest, args, system_function);
            ode_step_symplectic(var, var_copy, t, dt_s, w, local_gamma, system_dim, 
                                order, lowest, args, system_function);
        }
        else if (order == 2)
        {
            ode_step_symplectic_lowest(var, var_copy, t, dt_s, w, system_dim, 
                                       args, system_function);
            ode_step_symplectic_lowest(var, var_copy, t, dt_m, w, system_dim, 
                                       args, system_function);
            ode_step_symplectic_lowest(var, var_copy, t, dt_s, w, system_dim, 
                                       args, system_function);
        }
    }
}

void ode_step_symplectic_lowest(double var[], double var_copy[], double &t, 
                                double dt, double w, unsigned int system_dim, 
                                double *args, int (*system_function)(double t, 
                                const double s[], double f[], void *args))
{
    double step = 0.5 * dt;
    double f[system_dim];
    double f_copy[system_dim];

    /** Evaluate derivatives for Pa **/
    system_function(0.0, var, f, args);
    system_function(0.0, var_copy, f_copy, args);
    /** Apply Pa **/
    phi_a(var, var_copy, f, f_copy, step, system_dim);

    /** Evaluate derivatives for Pb **/
    system_function(0.0, var, f, args);
    system_function(0.0, var_copy, f_copy, args);
    /** Apply Pb **/
    phi_b(var, var_copy, f, f_copy, step, system_dim);
    
    /** Apply Pc **/
    phi_c(var, var_copy, dt, w, system_dim);
    
    /** Evaluate derivatives for Pb **/
    system_function(0.0, var, f, args);
    system_function(0.0, var_copy, f_copy, args);
    /** Apply Pb **/
    phi_b(var, var_copy, f, f_copy, step, system_dim);

    /** Evaluate derivates for Pa **/
    system_function(0.0, var, f, args);
    system_function(0.0, var_copy, f_copy, args);
    phi_a(var, var_copy, f, f_copy, step, system_dim);

    t += dt;
}

void phi_a(double var[], double var_copy[], const double function[], 
           const double function_copy[], double step, unsigned int system_dim)
{
    unsigned int n = system_dim / 2;
    
    for (unsigned int k = 0; k < n; k++)
    {
        var[k+n] += step * function[k+n];
        var_copy[k] += step * function_copy[k];
    }
}

void phi_b(double var[], double var_copy[], const double function[], 
           const double function_copy[], double step, unsigned int system_dim)
{
    unsigned int n = system_dim / 2;
    for (unsigned int k = 0; k < n; k++)
    {
        var[k] += step * function[k];
        var_copy[k+n] += step * function_copy[k+n];
    }
}

void phi_c(double var[], double var_copy[], double step, double w, 
           unsigned int system_dim)
{
    int n = system_dim / 2;
    double cwd = cos(2.0 * w * step);
    double swd = sin(2.0 * w * step);
    double tmp_var[system_dim];
    double tmp_var_copy[system_dim];

    for (int k = 0; k < n; k++)
    {
        tmp_var[k] = 0.5 * (var[k] + var_copy[k] + 
                            cwd * (var[k] - var_copy[k]) + 
                            swd * (var[k+n] - var_copy[k+n]));
        tmp_var[k+n] = 0.5 * (var[k+n] + var_copy[k+n] - 
                              swd * (var[k] - var_copy[k]) + 
                              cwd * (var[k+n] - var_copy[k+n]));
        tmp_var_copy[k] = 0.5 * (var[k] + var_copy[k] - 
                                 cwd * (var[k] - var_copy[k]) - 
                                 swd * (var[k+n] - var_copy[k+n]));
        tmp_var_copy[k+n] = 0.5 * (var[k+n] + var_copy[k+n] + 
                                   swd * (var[k] - var_copy[k]) - 
                                   cwd * (var[k+n] - var_copy[k+n]));
    }
    
    for (int k = 0; k < system_dim; k++)
    {
        var[k] = tmp_var[k];
        var_copy[k] = tmp_var_copy[k];
    }
}
\end{verbatim}

\subsection*{Equations of motion}

Despite being simple to obtain, the equations of motion used for the tests are made available in 
order to allow the faithful reproduction of the results shown here.

\begin{verbatim}
int system_function_2d_optical_lattice(double t, const double s[], 
                                       double f[], void *args) 
{
    (void)(t); // Avoid unused parameter warning
    double *par = (double *)args;
    double a = par[0];
    double U = par[1];
    
    f[0] = 2.0 * s[2];
    f[1] = 2.0 * s[3];
    f[2] = 2.0 * U * sin(s[0]) * (cos(s[0]) + a * cos(s[1]));
    f[3] = 2.0 * U * sin(s[1]) * (cos(s[1]) + a * cos(s[0]));
    
    return 0;
}

int system_function_three_body(double t, const double s[], 
                               double f[], void *args)
{
    (void)(t); // Avoid unused parameter warning
    double *par = (double *)args;
    double  n   = par[0];
    double mu_1 = par[1];
    double mu_2 = par[2];

    f[0] = s[2] + s[1];
    f[1] = s[3] - s[0];
    double r1_3rd = pow(sqrt((s[0] + mu_2) * (s[0] + mu_2) + s[1] * s[1]), 3);
    double r2_3rd = pow(sqrt((s[0] - mu_1) * (s[0] - mu_1) + s[1] * s[1]), 3);
    f[2] =  s[3] - s[0] + n * n * s[0] - mu_1 * (s[0] + mu_2) / r1_3rd - 
                                         mu_2 * (s[0] - mu_1) / r2_3rd;
    f[3] = -s[2] - s[1] + n * n * s[1] - (mu_1 / r1_3rd + mu_2 / r2_3rd) * s[1];

    return 0;
}
\end{verbatim}

Full code available in: \url{https://github.com/matheuslazarotto/SymplecticTao}.

\end{document}